# Proposal of a FET-LET Hybrid 6T SRAM

Antardipan Pal[1], Yong Zhang[1,*], *Senior Member, IEEE*, and Dennis D. Yau[2]

[1] Department of Electrical and Computer Engineering, University of North Carolina at Charlotte, Charlotte, NC 28223, USA

[2] Cupertino, CA 95014, USA

*Abstract*—There is an extremely high demand for a high speed, low power, low leakage, and low noise Static Random-Access Memory (SRAM) for high performance cache memories. The energy efficiency of SRAM is of paramount importance in both high performance and ultralow-power portable, battery operated electronic systems. In this article the factors affecting the overall speed and total energy consumption of a conventional 6T SRAM cell/array with 6 FETs, particularly roles of access transistors are analyzed to highlight the needs and directions for improvement. A hybrid 6T SRAM with two access FETs being replaced by light-effect transistors (LETs) and the electrical word lines replaced by optical waveguides (OWGs) is proposed. This hybrid SRAM is analyzed to reveal its potential in improvement of the switching speed and thus total energy consumption over the conventional 6T SRAM. Numerical analyses of a prototype hybrid SRAM array of 64 KB show a factor of 4 and 22 reduction in read delay and read energy consumption, respectively; and 3 and 4 in write delay and write energy consumption, respectively, when the access FETs are replaced by LETs. The potential impacts on the peripheral and assist circuits due to this hybrid structure and application of the LETs there are also briefly discussed.

* yong.zhang@uncc.edu

## I. Introduction

INTEGRATING electronic and photonic systems on the same chip can potentially transform computing architectures and enable more powerful computers. It is now possible to integrate a large number of electronic devices and photonic components on a single chip to perform logic, memory, and interconnect functions [1]. However, typically, the photonic components play only the roles of providing high-speed communications between different electronic subsystems [2], [3], for instance, in photonic DRAM [4], rather than any active roles in processing or modifying information like logic gates. Although photo-conductive devices can potentially offer advantages in switching speed [5] and switching energy [6], one major drawback of using such devices, like a light effect transistors (LETs) [6], [7], is the inconvenience of using the output of one LET based logic gate to directly drive the next similar logic gate without going through relatively inefficient electrical to optical energy conversion. To explore the advantages of LETs but avoid the cascading issue in computing applications [8], we seek to replace some field effect transistors (FETs) that only serve the roles of switching a circuit on and off, such as, the access transistors in a SRAM cell.

One of the most crucial concerns in many ultralow-power applications is energy efficiency. SRAM being one of the most critical building blocks in almost all digital systems, its packing density, speed, power consumption are all crucial performance metrics, [9]. SRAMs are generally used in high speed cache memories providing a direct interface with a CPU at high speeds which are not possible to attain by other memory circuits. However, on chip caches typically consume 25%–45% of the total energy of a chip [10]. Moreover, in modern high-performance large density memory circuits, more than 40% of the total energy is consumed due to leakage currents [11].

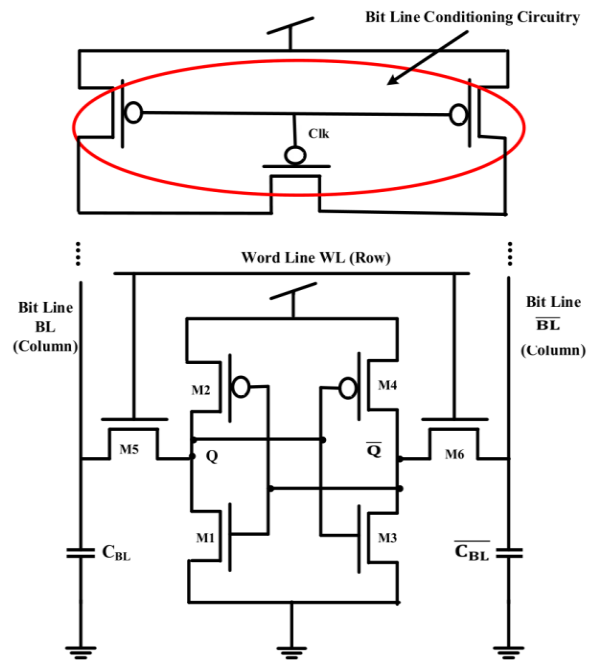

Fig. 1. A 6T SRAM Cell.

Hence, high-speed and energy-efficient embedded memories are desirable for a modern electronic system.

A conventional 6T SRAM cell is shown schematically in Fig. 1, consisting of 6 FETs labeled as M1-M6. (M1&M2) and (M3&M4) form the cross-coupled inverter pairs (latches), and M5 and M6 are the access devices that allow the data stored in the cell to be accessed and modified by charging and discharging the output nodes Q and $\overline{Q}$ and bit lines BL and $\overline{BL}$ during the read and write operations. The two access transistors play an extremely crucial role in determining the overall speed, power dissipation and stability of the cell [12], [13]. Additionally, the three p-FETs, encircled in red in Fig. 1, are the bit line conditioning devices whose roles are to pre-charge and equalize the bit line voltages before each read and write operation.

Their switching speed and energy consumption are also critical to the performance of the SRAM [13], [14]. Many approaches in both device and circuit level have been explored to offer various incremental improvements in the SRAM performance, particularly in speed and energy consumption [15-17].

Our proposal of a FET-LET hybrid 6T SRAM technology represents a drastically different approach that can offer major improvement on the read and write speeds and the corresponding energy consumptions by replacing the two access FETs with two LETs and accordingly the word line electrical wires with optical waveguides (OWGs). This idea offers a more intimate integration of the electronics and photonics, namely on the CPU chip level. Additionally, this application avoids the well-known energy-data rate (EDR) challenge (EDR ≤ 10fJ/bit for on-chip communication) [2], [18] because it does not require using light to address photonic devices individually [8], but in a group simultaneously through an optical waveguide. The focus of this work is on the potential improvement of the 6T SRAM array itself, but the anticipated benefits of applying LETs in the peripheral circuits will also be discussed briefly.

## II. ANALYSIS OF DELAY AND ENERGY CONSUMPTIONS IN A 6T SRAM ARRAY

The primary factors limiting the read and write speeds and the corresponding energy consumptions of a 6T SRAM array are the capacitances of the bit lines and word lines, and the characteristics of the access transistors.

### A. Read and Write Delay Calculations

The critical capacitances of a 6T SRAM cell/array are calculated as follows [13], [19]:

$$C_{BL} = n_R(C_{drain\_access} + C_{ht}) \quad (1)$$

$$C_{WL} = n_C(2C_{gate\_access} + C_{width}) \quad (2)$$

$$C_{out} = C_Q = C_{drain,M1} + C_{drain,M2} + C_{gate,M3} + C_{gate,M4} + C_{drain,M5} \quad (3)$$

where $C_{BL}$ is the bit line capacitance, $C_{WL}$ is the word line capacitance, $C_{out}$ is the capacitance of the output node, $C_{drain\_access}$ and $C_{gate\_access}$ are the drain and gate capacitances respectively of the access devices, $n_R$ and $n_C$ are the number of rows and columns respectively for the SRAM array. The bit line wire capacitance ($C_{ht}$) and the word line wire capacitance ($C_{width}$) are calculated using relations from [13] and parameters from [20], [21]. All the drain and gate capacitances of the access FETs (M5,M6) and the core FETs (M1-M4) are calculated using relations from [12], [22] and FET model parameters are from [23].

The read and write delays are calculated as follows [13]:

$$T_{read} = \frac{C_{WL}V_{DD}}{I_{word\_drive}} + \frac{C_{BL}\Delta V_{read}}{I_{read}} \quad (4)$$

$$T_{write} = max\left(\frac{C_{WL}V_{WL}}{I_{WL}}, \frac{C_{BL}V_{DD}}{I_{write\_ckt}}\right) + \frac{C_{out}\Delta V_{out}}{I_{write}} \quad (5)$$

where $I_{read}$ and $I_{write}$ are the 6T cell read and write currents respectively; $I_{word\_drive}$, $I_{WL}$, and $I_{write\_ckt}$ are the word line driver, word line, and write circuitry currents respectively [13]; $\Delta V_{read}$, and $\Delta V_{out}$ are respectively the change in the bit line voltage after read operation and the change of output voltage after write, and $V_{DD}$ is the supply voltage.

### B. Read and Write Energy Calculations

Read and write energies are calculated as follows [13]:

$$E_{read} = C_{WL}V_{DD}^2 + C_{BL}V_{DD}\Delta V_{read} \quad (6)$$

$$E_{write} = C_{WL}V_{DD}V_{WL} + C_{BL}V_{DD}^2 + C_{out}V_{DD}\Delta V_{out} \quad (7)$$

where $V_{WL} = V_{DD}$ is the word line voltage. Note that for the above delay and energy equations of the 6T array, the delay and switching energy due to the transit of carriers through the FET channel has not been considered, since they are negligible compared to the gate related RC delays and energies.

As evident from the above formulas, the access transistors and the ways to address them play critical roles in determining the overall SRAM cell performance. Therefore, if the access devices can be replaced with some high-speed switching devices with very low gate, source and drain junction capacitances, such as LETs, as described in the next section, and also can be addressed more efficiently, major improvement in speed and power consumption can be achieved.

### C. Energy Consumption Associated with Leakage Currents

There are various types of leakage currents in a modern FET that contribute to the energy loss. They include subthreshold leakage current $I_{sub}$, gate leakage $I_{gate}$ (gate-induced drain and source leakage current, gate tunneling leakage current through the bulk, source, and drain), and junction leakage current $I_{junction}$ (p-n junction leakage currents at the drain-substrate and source-substrate junctions) [24], [25]. Although in the static state, the leakage currents of the two inverters dominate the static energy consumption, during the write and read processes, the leakage of access transistors M5 and M6 also contributes to the total energy consumption. From [25], it can be roughly estimated that about 40% of the total leakage in a 6T cell is in the access paths.

## III. LIGHT EFFECT TRANSISTOR (LET)

### A. Device Overview and Advantages over FET

A LET as shown in Fig. 2 is a semiconductor nanowire (SNW) placed on an insulating substrate with two metal contacts at the ends [6]. Working mechanism of a LET is different from that of a traditional FET in that the source-drain conductivity of a LET is modulated by light or EM radiation of a suitable wavelength as in photoconductive mechanism [6], [26]. The advantage of a LET over a FET stems from various factors like removal of physical gate, thus minimizing the complex gate fabrication process and random dopant fluctuations in FETs [27]. Hence, the LET can be scaled down to quantum regime without the problem of short-channel effects (SCEs) that are common in nanoscale FETs [28].

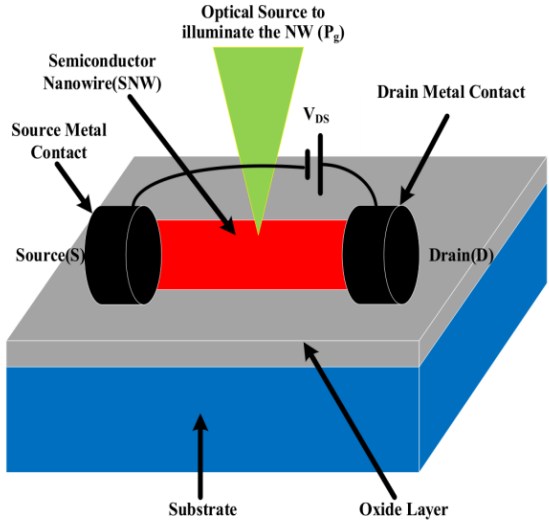

Fig. 2. A Light Effect Transistor (LET).

Also, because the LET structure does not have a physical gate, the device speed is expected to be only limited by the carrier transit time or lifetime, whichever is smaller, rather than the capacitive delay as in the gated FET.

Although the demonstrated prototype LETs were based on CdSe NWs [6], there is no limit to the material system per the device mechanism. At room temperature, many semiconductors (e.g., Si, SiC, InAs, InP, GaAs, CdSe) have saturation electron velocities in the range of (1-10) x $10^7$ cm/s when the electrical field is on the order of 100 kV/cm [29], [30], which implies a carrier transit time of the order of (1-0.1) ps for a 100 nm long NW. 100 nm is also the typical length scale of ballistic transport where the saturation velocity can be achieved. For longer NWs in the non-ballistic transport regime, the electron transit time depends on the electrical field. For Si at $E$ = 10 kV/cm, the electron velocity is around 7 x $10^6$ cm/s [29], [30], and the carrier transit time ($t_{LET}$) can be estimated to be 4.3 ps for a 300 nm long Si NW. If the NW in the LET can be scaled down to operate in the ballistic regime (typically ≤ 100 nm), then ultra-fast switching (of the order of 1 ps or faster) can be obtained. The ultra-fast switching of the LET translates to ultra-small switching energy. For instance, assuming a switching time (carrier transit time) $t_{LET}$ = 1 ps (for a ballistic device), an on-current of $I_{sd}$ = 1 µA under $V_{sd}$ = 1 V, the electrical switching energy $E_{el}$ (= $I_{sd}$ x $V_{sd}$ x $t_{LET}$) will be of the order 1 aJ/switch. However, in the LET, optical gating power also contributes to the switching energy. The net gating power required can be estimated by $P_g = E_{ph}I_{sd}/(eG)$, where $E_{ph}$ is the photon energy and $G$ is the photo-conductive gain. Assuming $E_{ph}$ = 2.5 eV, $G$ = $10^3$, to have $I_{sd}$ = 1 µA, we get $P_g$ = 2.5 nW. Then, assuming $t_{LET}$ = 1 ps, the optical switching energy $E_{op}$ will be 2.5 x $10^{-3}$ aJ/switch << $E_{el}$, which leaves sufficient room allowing for below 100% light power delivery efficiency. In an even more idealistic case, assuming a ballistic device with a quantum impedance of 12.9 kΩ [31], transit time of 0.1 ps, S-D current of 1 µA, and no voltage loss at the contacts, the electrical switching energy can be as low as 1.3 x $10^{-21}$ J/switch at a very low $V_{sd}$ of only 13 mV [6]. For a prototype device, a 5 µm long and 80 nm in diameter CdSe NW LET structure studied previously [6], under 532 nm illumination of 110 nW (only about 6% was actually absorbed), yielded $I_{ds}$ = 0.35 µA at $V_{ds}$ = 1.43 V; in dark, $I_{ds}$ ~ 1 pA, which corresponds to 1.5 pW static or off power. Estimating the switching energy for such a large device using the typical room temperature carrier lifetime in a II-VI semiconductor in the order of 100 ps, the total switching energy $E_{tot,sw} = E_{el} + E_{op} \approx 0.06$ fJ/switch would still be better than typical FETs having switching energy of 0.1-1 fJ/switch [32]. In FETs, the gate related RC delays predominate over the transit-time delay; but in the LET, the carrier transit time through the NW channel is expected to be the predominant factor for determining the switching speed and energy of a discrete LET. Moreover, the $I_{on}/I_{off}$ ratio for a LET could be as high as $10^6$ [6], which is almost an order of magnitude better than that of advanced FETs. This reduces the leakage in the access paths and offer more flexibility in the 6T cell design with LET access devices.

*B. Hybrid 6T SRAM with Access FETs Replaced by LETs*

To take advantage of the high switching speed and low energy consumption of LETs, the two access transistors (M5 and M6) in the 6T cell of Fig. 1 are replaced by two LETs (L1 and L2) as shown in the prototype hybrid 6T cell of Fig. 3, where the word line is replaced by an optical waveguide (OWG) that transmits light to the LETs.

To quantify the potential improvement, we consider a design with a moderate size LET based on a generic semiconductor NW: L = 300 nm (length) and D = 50 nm (diameter) and supported on an insulating substrate (e.g., SiO$_2$/Si) as shown in Fig. 2. Also, a ballistic device with smaller dimensions (L = 100 nm and D = 30 nm) is considered which yields a much reduced 6T cell and array area and high cell density.

For the LET structure, there will be no MS-junction capacitance that is equivalent to the drain or source capacitance (gate-drain or gate-source overlap capacitance along with the drain-substrate or source-substrate junction capacitance) of FETs, since there is neither a gate nor any electrical paths to ground between the MS structure and the substrate that has no electrical connection as opposed to the doped substrates of FETs. The photocurrents of the NW photodetector are typically in the range of 1-10 µA [33-35]. For the LET access device, the on-current is assumed to be 5 µA. The switching delay, as estimated by the transit time earlier, is assumed to be around 4 ps and 0.1 ps for the non-ballistic and ballistic cases, respectively.

*C. Critical Capacitance, Read, Write Delay and Energy Consumption of the 6T SRAM with LET Access Devices*

The critical capacitances of the hybrid 6T SRAM with LET access devices are modified from the $C_{BL}$ and $C_{out}$ calculated in (1) and (3) as follows:

$$C'_{BL} = n_R(C_{ht}) \quad (1')$$

$$C'_{out} = C_Q = C_{drain,M1} + C_{drain,M2} + C_{gate,M3} + C_{gate,M4} \quad (3')$$

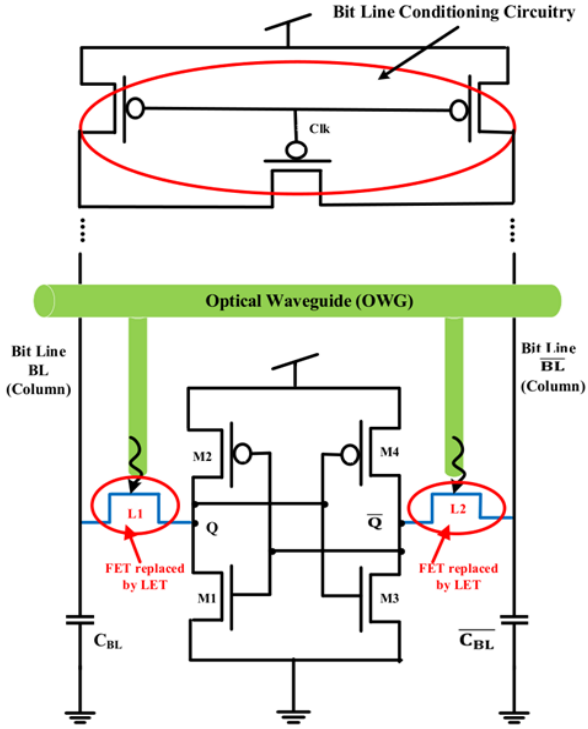

Fig. 3. A prototype hybrid 6T SRAM cell with LET access.

In (1′), the modified bit line capacitance is predominantly the wire capacitance only, since the LET access devices do not have any MS junction capacitance. In (3′), there is only the drain and gate capacitances of the core FETs. The drain equivalent capacitance of the access FET ($C_{drain,M5}$ in (3)) is not present in case of LET access devices due to the same reason. In the LET accessed SRAM, the word-line capacitance ($C_{WL}$) in (2) should be practically zero, since the access LETs neither have any gate capacitance, nor require a wired electrical signal to control the gates as in the case of access FETs.

Accordingly, the read and write delay and the corresponding energies in (4)–(7), are modified as below, with $I_{read}$ and $I_{write}$ being replaced by $I'_{read}$ and $I'_{write}$ appropriate for the LET access devices, and all the $C_{BL}$ and $C_{out}$ are replaced by $C'_{BL}$ and $C'_{out}$, keeping other parameters almost unchanged.

$$T_{read,LET\_access} = t_{WG} + max\left(\frac{C'_{BL}\Delta V_{read}}{I'_{read}}, t_{LET}\right) \quad (4')$$

where the RC-word-line delay during read operation in (4) is replaced by the time taken by the optical signal to propagate through the optical wave-guide ($t_{WG}$), and the second term is the larger term of the modified RC-bit line delay during read and the LET carrier transit delay.

$$T_{write,LET\_access} = max\left(t_{WG}, \frac{C'_{BL}V_{DD}}{I'_{write\_ckt}}\right) +$$
$$max\left(\frac{C'_{out}\Delta V_{out}}{I'_{write}}, t_{LET}\right) \quad (5')$$

where the first term in (5′) is the larger term of $t_{WG}$ and the modified bit line delay during write, and the second term is the larger term of the modified 6T cell flipping delay during write and the LET carrier transit delay.

$$E_{read,LET\_access} = (2n_c E_{op}) + max\left(C'_{BL}V_{DD}\Delta V_{read}, E_{el}\right) \quad (6')$$

$$E_{write,LET\_access} = (2n_c E_{op}) + C'_{BL}V_{DD}^2 + max(C'_{out}V_{DD}\Delta V_{out}, E_{el}) \quad (7')$$

where the word line energies in (6) & (7) (the first terms) will not be present in case of LET accessed cells or arrays, assuming the light propagation loss through OWG is practically negligible. The first terms in both (6′) and (7′) are the optical gating switching energy ($E_{op}$), and for a whole row it is multiplied by $2n_C$ where $n_C$ is number of 6T cell in a row and each 6T cell has 2 LET access devices. The second term in (6′) is the larger of the modified bit line RC-read energy, and the LET carrier transit electrical switching energy ($E_{el}$) as described in Section III-A, while the last term in (7′) is the larger term of the modified cell flipping RC-energy during write, and the LET carrier transit electrical switching energy.

Therefore, it can be expected that a SRAM array with LETs in the access paths will reduce all the critical capacitances (except for the bit line wires) as compared to the array with FET access devices. Also, the carrier transit delay and switching energy (depending on the transit delay) of a LET are much lesser as compared to the RC delay and switching energy of a FET.

Analytical relations (4)-(7) and (4′)-(7′) are used to calculate a set of delays and energy consumptions for various 6T- SRAM arrays (32 bytes - 64 KB) with respectively, FET, LET and ballistic LET access devices for direct comparison. Despite as approximations, these analytical relations offer more transparent insight to the underlying device physics than numerical simulations. To ensure their ability to yield results of acceptable accuracy for the purpose of comparing two vastly different technologies, the analytical relations, (4)-(7), were first used to calculate read and write delays following simulation conditions used in [36], which could reproduce the simulation results of [36] with 26% average accuracy. The values of the currents in the 6T FET SRAM are assumed to be 25 µA, considering the effective drive currents of 22 nm FETs [37] and $\Delta V_{read} \approx 120$ mV and $\Delta V_{out} \approx V_{DD}/2$ [13]. The results for different SRAM array sizes are shown in Figs. 4(a)-(d), and the numerical results are given in Table I for 4-KB and 16-KB arrays. It is clear from the results, as summarized in Table I, that using L ET access devices may result in marked improvement in the overall delay and energy consumption of the SRAM array. From the delay and energy plots of Fig. 4, it is found that the results for LET and ballistic LET are coinciding despite the ballistic device having much lesser carrier transit delay and switching energy as compared to non-ballistic LETs. This is because for an array, the overall RC delay and energy will dominate over the carrier transit delay and switching energy of the individual LETs. On read delay, Fig. 4(a) reveals approximately a factor of 4 average reduction with LET access devices over FET access devices. Accordingly, on read energy, Fig. 4(b) reflects approximately a factor of 18 average reduction. On write delay, the average reductions shown in Fig. 4(c) is approximately a

TABLE I

COMPARISON OF THE PERFORMANCE OF 4-KB AND 16-KB SRAM ARRAYS WITH FET, LET, AND BALLISTIC LET ACCESS DEVICES

|  | 4-KB SRAM Array | | | 16-KB SRAM Array | | |
| --- | --- | --- | --- | --- | --- | --- |
|  | **FET Access Devices** | **LET Access Devices** | **Ballistic LET Access Devices** | **FET Access Devices** | **LET Access Devices** | **Ballistic LET Access Devices** |
| **Read Delay (ps)** | 846 | 187.1 | 187 | 1692 | 374.2 | 373.9 |
| **Write Delay (ps)** | 790.4 | 302.6 | 302.6 | 1576.8 | 598.3 | 598.3 |
| **Read Energy (fJ)** | 20.1 | 0.9 | 0.89 | 40.2 | 1.8 | 1.77 |
| **Write Energy (fJ)** | 30 | 7.07 | 7.06 | 59.8 | 14.1 | 14.1 |

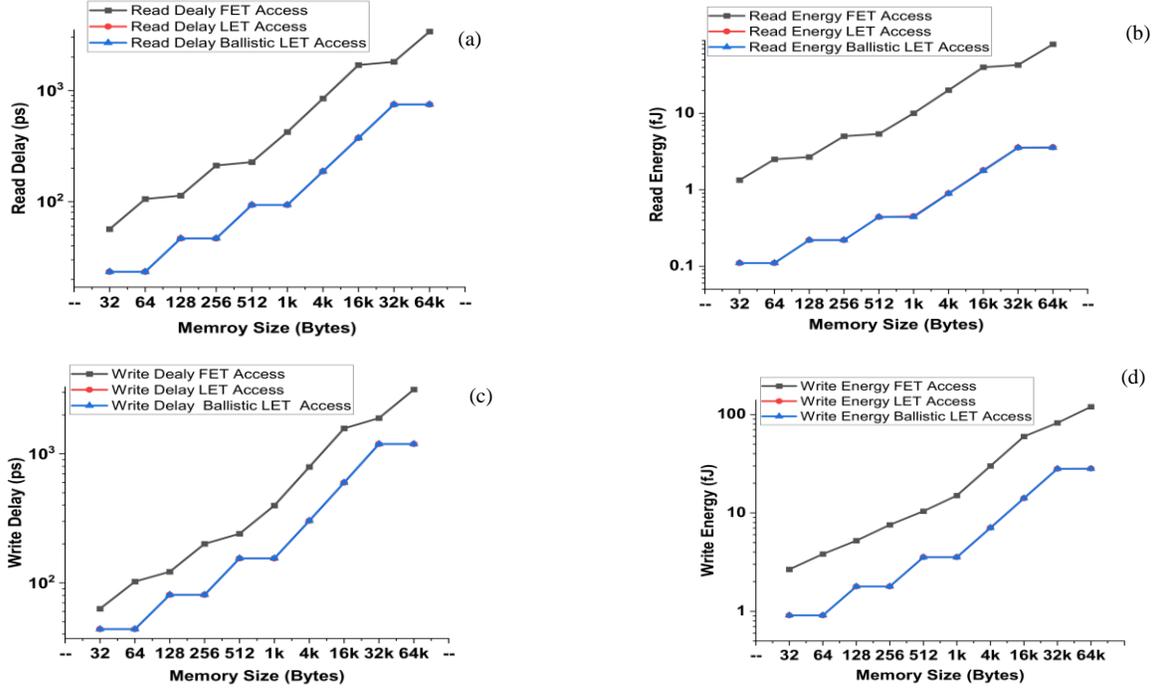

Fig. 4. Read and write delay and energy for various SRAM arrays with FET, LET and ballistic LET access devices. (a) Read Delay. (b) Read Energy. (c) Write Delay. (d) Write Energy. The curves with LETs and ballistic LETs are indistinguishable.

factor of 2, and on write energy, Fig. 4(d) shows approximately a factor of 4 average reduction. The results for both the read and write energy can indeed satisfy the requirement of EDR ≤10 fJ/bit for on-chip photonic integration [2], [3] as shown in Table I. The general operating principle, mechanism, and conclusion are in-principle applicable for LETs appropriately fabricated with any semiconductor

### D. Discussion on the Delay and Energy Consumption of the 6T Array with FET and LET Access Devices

*6T Array without Peripherals*

The improvement in the read delay of the hybrid 6T SRAM array is primarily due to the removal of the word line delay in (4) by the optical-waveguide delay ($t_{WG}$) in (4′) which is almost negligible as compared to the RC word line delay with FET access devices. The improvement in the write delay is due to the removal of the $C_{WL}$ related term and reduced overall bit line capacitance from $C_{BL}$ to $C'_{BL}$. Relatively, the improvement in the write delay is lesser than the read delay, because the first term in the write delay in (5) takes the larger one of the two contributions, and hence the advantage of replacing the word line delay in (5′) by $t_{WG}$ does not affect the overall write delay as much as it affects the read delay. Also, the highest reduction achieved in the read energy is mainly due to the replacement of the RC word line energy consumption in (6) (the first term) with FET access devices by a much smaller optical gating term in (6′) (the first term) with LET access devices. Similarly, the improvement in the write energy in (7′) is lesser than read energy in (6′), due to the presence of the second and third terms in (7′) where $V_{DD} > \Delta V_{out} > \Delta V_{read}$, and thus $C'_{BL} V_{DD}^2$, the second term in (7′) $> max (C'_{BL} V_{DD} \Delta V_{read}, E_{el})$, the second term in (6′), which to some extent lessens the amount of improvement in the write energy as compared to read energy .

Note that in above analyses, the hybrid 6T SRAM array offers improved performance even with a smaller drive-current (by a factor of 5) than the conventional 6T SRAM array. Since much lesser current is needed in the 6T cell with LET access devices, the core FETs (the FETs in the inverter pairs) can be scaled down to lesser device widths (FET drive current is proportional to the device width), which may offer saving in area for the 6T cell and hence for the whole array. However, we would like to point out that if the on-current of the hybrid 6T SRAM is increased to the same level as the conventional 6T SRAM, the read delay can be further reduced by approximately a factor of 5, but the improvement on write delay may be less significant, because for a large array it primarily depends on $I_{write\_ckt}$.

The figure of merit (FOM) of the SRAM array can be found from the energy-delay product (EDP) [13]. Considering 50% probability of the array being accessed in a cycle, and 50% probability for each of the read and write operations [13], it can be roughly estimated that the hybrid SRAM arrays (average EDP of the order of $10^{-24}$ Js) on an average exhibit more than one order of magnitude lesser EDP, as compared to regular SRAM arrays (average EDP of the order of $10^{-23}$ Js).

*Possible Improvements in the Peripheral and Assist Circuits*

Till now we have only focused on the effects of LET access devices on the 6T array and have not considered the potential benefits on the peripheral circuits [13], [38] either as a direct consequence of replacing the access devices in the SRAM cells, or even further by replacing some FETs with LETs in the peripheral or assist circuits. Firstly, replacing the FET access devices with the LETs in the 6T SRAM, and thus replacing the electrical word-lines with OWGs abolishes the need of the word line drivers, which not only reduces the capacitance and the RC-electrical energy consumption of the word line to practically zero, but may also reduce the capacitance and energy consumption of the column decoder circuitry [13]. It can be roughly estimated that over various array sizes, the hybrid array on an average has almost two and three orders of magnitude lesser word line delay and energy consumption respectively over the regular array with word line drivers [13]. A prototype 6T array with LET access devices and OWG, keeping the core FETs and other peripherals almost unchanged is shown in Fig. 5 below. The electrical row decoder circuit in a conventional SRAM array has to be replaced by an opto-electronic counterpart to illuminate the OWGs by appropriate optical sources, for example, nanoscale lasers [18].

Secondly, besides using LETs as the access devices, there may be a scope to replace some FET based switches in other peripherals and assisting circuitry of the 6T array by LETs, which will further reduce the relevant delay and energy consumption, and hence further improve the performance of the SRAM array. For instance, there may be a possibility to replace the three p-FETs of the bit line conditioning circuitry of Fig. 3 by LETs, which will further reduce the bit line capacitances and hence bit line related delays and energy consumptions, especially for large 6T arrays.

*E. Improvement on Leakage Using LET Access Devices*

For the LET structure, all the leakage mechanisms (currents) for the FET mentioned in Section II-C are eliminated except for the subthreshold current that is equivalent to the dark current of the LET. Since generally doping is not required for the LET, it can have very low dark current (e.g., of order of few pA) [6].

LETs have a different turn on mechanism and no SCEs as discussed previously [6] and hence hybrid SRAMS will have minimal subthreshold leakage in the access paths. Since LETs do not have a physical gate, hence there will be neither any gate related nor any SCE induced leakage [24], [25] in the access paths, and thus the leakage power consumption in the hybrid SRAM will be much reduced. Also, LETs do not have any p-n junctions or leakage paths to ground, and hence the hybrid SRAM will also have no junction leakage [24], [25] in the access paths and hence the overall leakage will be much reduced. It can be estimated from [25], that there will be an overall reduction of roughly 35% in the total leakage current in a single hybrid 6Tcell, which will be more advantageous in the case of a hybrid 6T array having a large number of such 6T cells.

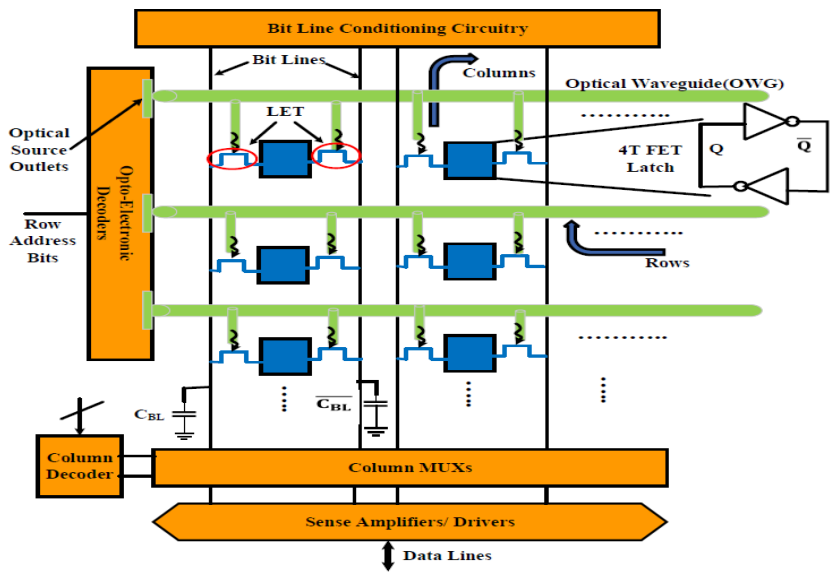

Fig. 5. Prototype of a LET accessed hybrid 6T SRAM Array

*F. Compatibility with MOSFET and Scalability*

In a typical circuit layout, the word-line spacing is "2 poly pitches", which is of the order of 200 nm for 22 nm technology node and 100 nm for 7 nm technology node [39], [40], respectively. OWGs fabricated on an insulating substrate can achieve subwavelength lateral size with very low loss, for instance, a Si waveguide of 400 nm for light at 1.55 μm with only 2.8 dB/cm loss [41]. Since photonic properties are scalable with wavelength, for the LETs operating in visible wavelengths, the OWG dimension can be significantly reduced (e.g., to around 140 nm at 532 nm). Further reduction is possible by using plasmonic-dielectric hybrid waveguides, although with somewhat higher loss [42]. Alternatively, to alleviate the size mismatch between the light wavelength and the electronic device, one may envision the possibility to adopt somewhat different circuit layouts in the hybrid SRAM (e.g., arranging multiple cells of the same word line together such that they can be optically addressed as a group), which can not only use the optical energy more efficiently but also increase the word line spacing. Additionally, as we know, the primary reasons for going down to smaller technology nodes are not only for saving space but also for improving performance. For the situations where the latter goal is more important, one could simply allow for larger waveguide spacing.

The optical waveguides are typically designed for interconnection between the circuits. In the hybrid SRAM, the optical paths are substantially shorter for on chip operation. Thus, the scalability to a few hundred or even over one thousand cells/WL is not expected to be an issue. The minimum light power output required from the optical decoder can be estimated to be in the order of only 8.5 μW per OWG for a 1600 cells/WL, considering a propagation loss of 2.8 dB/cm [41] and an overestimated OWG length of 1 mm, and using the optical gating power estimate given in Section III-A, which leaves a large room for less efficient implementation.

Heterogeneous integration of Si electronics with electronic and photonic components/structures made from compound semiconductors and other dielectric materials has been reported extensively with CMOS compatible process flows [42-44]. The necessary technologies have largely been demonstrated for different applications, for instance, in the hybrid InGaAs/SiGe 6T SRAM [45], and LiNbO3 photonic waveguide cavity on Si [46], which all can be readily transferred to the proposed new integration scheme.

## IV. Conclusion

LETs offer high speed and low energy opto-electronic switching, where the switching delay is limited by carrier transit time, which can be made extremely small by using the nanowire based device, particularly in the ballistic transport mode. In contrast, in the FETs, it is generally RC switching, and hence it is much less energy efficient due to high gate-related capacitances in FETs. The biggest advantage of replacing the FET access devices by LETs is that the gate, source and drain related capacitance and electrical word line are no longer present, which removes the word line delay as well as energy consumption. From the above delay and energy calculations of the hybrid 6T SRAM using LET access transistors, it can be concluded that this new hybrid 6T cell and array is much more energy efficient with lesser read and write delays as compared to the all FET 6T cells and arrays. In addition, LETs are expected to have much lower leakage currents than conventional FETs, and thus the hybrid 6T cell and array will have much lesser leakage power dissipation compared to those with FET access devices. The use of the optical waveguide-based word line architecture in the hybrid SRAM array abolishes the need of electrical word lines and also the word line drivers, which drastically reduces the total word line capacitance, RC-delay and energy consumption to almost negligible compared to that in the conventional SRAM array. Furthermore, LETs may find useful applications in other peripheral and assist circuits of the SRAM array like the bit line conditioning circuit for improvement in speed and energy consumption. The proposed hybrid SRAM architecture offers an example of hybrid electronic-photonic integrated circuit with both electronic and photonic devices playing active roles synergistically.

### Acknowledgment

This work was partially supported by fund from Bissell Distinguished Professorship at UNC-Charlotte. We thank Drs. Qiuyi Ye, Kwok Ng, and Amrutur Bharadwaj for valuable discussions and suggestions.